\def\removeHeaders{yes}

\def\tempYes{yes}

\documentclass[sigconf, 10pt, letterpaper,dvipsnames\ifx\removeHeaders\tempYes ,nonacm\fi]{acmart}

\renewcommand\footnotetextcopyrightpermission[1]{} 
\setcopyright{none}

\acmDOI{}

\acmISBN{}

\acmPrice{}

\acmYear{2019}
\copyrightyear{2019}
\acmConference[BS'19]{Buffer Sizing Workshop}{December 2-3}{Stanford, CA}

\usepackage[utf8]{inputenc}

\usepackage{enumitem}
\usepackage{algorithmicx}
\usepackage{algpseudocode}
\usepackage{algorithm}
\usepackage{graphicx}
\usepackage{subfig}

\usepackage[nomain, toc, acronym]{glossaries}
\glsdisablehyper

\makeatletter
\newcommand{\StateIndent}[1][3]{%
  \setlength\@tempdima{\algorithmicindent}%
  \Statex\hskip\dimexpr#1\@tempdima\relax}
\algdef{S}[WHILE]{WhileNoDo}[1]{\algorithmicwhile\ #1}%
\makeatother

\algnewcommand\True{\textbf{true}}
\algnewcommand\False{\textbf{false}}
\algnewcommand\Break{\textbf{break}}

\newcommand{\codel}{CoDel}

\begin{document}

\title{Cocoa: Congestion Control Aware Queuing}


\newacronym{aqm}{AQM}{Active Queue Management}
\newacronym{cc}{CC}{Congestion Control}
\newacronym{cca}{CCA}{Congestion Control Algorithm}
\newacronym{qdisc}{qdisc}{queuing discipline}
\newacronym{fq}{FQ}{Fair Queuing}
\newacronym{rtt}{RTT}{Round-Trip Time}
\newacronym{gi}{GI}{Guard Interval}
\newacronym{li}{LI}{Longest Interval}
\newacronym{bdp}{BDP}{Bandwidth Delay Product}
\newacronym{cocoa}{cocoa}{Congestion Control Aware qdisc}

\begin{abstract}
Recent model-based congestion control algorithms such as BBR use repeated measurements at the endpoint to build a model of the network connection and use it to achieve optimal throughput with low queuing delay. Conversely, applying this model-based approach to Active Queue Management (AQM) has so far received less attention. We propose the new AQM scheduler \textit{cocoa} based on fair queuing, which adapts the buffer size depending on the needs of each flow without requiring active participation from the endpoint. We implement this scheduler for the Linux kernel and show that it interacts well with the most common congestion control algorithms and can significantly increase throughput compared to fair \codel{} while avoiding overbuffering.
\end{abstract}

\author{Maximilian Bachl, Joachim Fabini, Tanja Zseby}
\affiliation{\institution{Technische Universität Wien}}
\email{firstname.lastname@tuwien.ac.at}



\settopmatter{printfolios=true}
\maketitle

\section{Introduction}

In the last decades various \gls{aqm} mechanisms have been proposed to minimize excessive \textit{standing queues} in the Internet. One of the most influential recent efforts is \codel{} \cite{nichols_controlling_2012} whose goal is that the queuing delay at the bottleneck link is at least once under 5\;ms in a moving window of 100\;ms. 

While it is important to keep the queuing delay constrained, it is also necessary to ensure that overly aggressive flows cannot benefit from ``stealing'' less aggressive flows' bandwidth. Thus researchers and engineers have developed \textit{\gls{fq}} mechanisms \cite{shreedhar_efficient_1996,dumazet_pkt_sched:_2013} to isolate different flows' queues so that for example a delay-sensitive live video call cannot be impaired by a concurrent bulk transfer which consumes all the available bandwidth. 

Recent approaches have tried to combine \gls{aqm} with \gls{fq}. \cite{taht_flow_2018} demonstrate \textit{fq\_codel}, a \gls{qdisc} that uses \gls{fq} and lets \codel{} manage each queue. \cite{hoiland-jorgensen_piece_2018} expand upon this and create the \textit{cake} \gls{qdisc} that also adds features such as not only per-flow queuing but also per-host queuing for even increased fairness. Furthermore they also include bandwidth shaping into their solution and aim to create one \gls{qdisc} that is easy to configure, can be easily deployed on home routers and offers all features in one solution. 

While we do not want to make statements about the general performance of \codel{}, we show that fq\_codel and cake do not optimally use available bandwidth in common network configurations for common \glspl{cca}. This becomes especially prevalent for links with a high bandwidth or a large \gls{rtt} but is already noticeable for common scenarios, such as a link with 100\;Mbit/s and an \gls{rtt} of 50\;ms. We show that the impaired performance is a result of keeping the queuing delay under 5\;ms, which hinders \glspl{cca} such as Reno or Cubic from reaching maximum throughput. Moreover, this behavior can result in unnecessary standing queues on links with very low latencies under 10\;ms, which occur in data centers and between users and close by servers of content delivery networks. 

As a remedy, we conceive a fair \gls{aqm} mechanism that explicitly measures the behavior of the \gls{cc} of a flow and dynamically changes the buffer so that 
\begin{enumerate}[topsep=0pt,wide,labelwidth=!,labelindent=0pt]
\item link utilization is maximized and
\item queuing delay is kept at the minimum that is required to achieve optimum throughput considering the \gls{cc}.
\end{enumerate}
The concept of dynamically adjusting the buffer size depending on flows' needs was recently proposed by \cite{bless_policy-oriented_2018}. However, their solution is not tailored for fair queuing but for flows sharing a queue. 

We show that a prototype of our solution \textit{\gls{cocoa}} can achieve the aforementioned objectives for the most common loss-based \glspl{cca}, Reno \cite {jacobson_congestion_1988} and Cubic \cite{ha_cubic:_2008}. These \glspl{cca} operate by continuously increasing the number of bytes that are allowed to be in the network (congestion window) if no packet loss is experienced and by sharply decreasing this number if a packet is lost (multiplicative decrease). With Cubic being the default \gls{cc} in all major OS, we especially emphasize our evaluation on improving its performance. Contrasting to the aforementioned \glspl{cca}, recently proposed BBR \cite{cardwell_bbr:_2016} does not continuously increase and then sharply decrease when packets are lost but instead uses periodic measurements to estimate the available bandwidth as well as the minimal \gls{rtt} and then tries to stay at this point of optimal bandwidth and minimal delay. BBR is thus considered to be \textit{model-based}. We show that \gls{cocoa} also behaves well in interaction with BBR~v1.

\section{Concept}

\begin{figure}[h]
\includegraphics[width=\columnwidth]{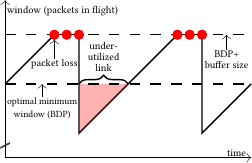}
\caption{If the buffer is too small, loss-based \glspl{cca} cannot fully utilize the link since they send too few data following the multiplicative decrease that occurs after packet loss.}
\label{fig:tooLittle}
\end{figure}

\begin{figure}[h]
\includegraphics[width=\columnwidth]{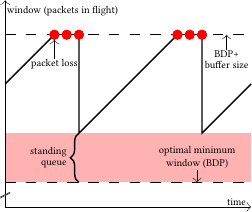}
\caption{If the buffer is too large, loss-based \glspl{cca} keep an unnecessary standing queue not required for achieving full link utilization.}
\label{fig:tooMuch}
\end{figure}

Since the goal of our algorithm is to achieve optimal throughput irrespective of the \gls{cca}, we aim to avoid the scenario depicted in \autoref{fig:tooLittle}: Here the buffer is too small, meaning that the \glspl{cca} never manages to achieve full utilization and periodically underutilizes the link, leading to, for example, a user waiting longer for a software update to finish or a game to download on their video game console. 

The other case we want to avoid is a persisting standing queue, as depicted in \autoref{fig:tooMuch}: In this case the buffer is too large meaning that optimal throughput is achieved but at the same time that an unnecessarily long standing queue is maintained. This oversized queue leads to \gls{rtt} being larger than necessary and can result in unresponsive applications and reduced Quality of Experience. Furthermore, the buffer space has to be allocated in the bottleneck device and keeps the memory from serving a more useful purpose. 

These considerations lead us to designing an algorithm that dynamically measures the \gls{cca} and adapts the buffer size to reach our goal of maximum throughput and minimal delay. The basic functionality of this algorithm would be to
\begin{enumerate}[topsep=0pt,wide,labelwidth=!,labelindent=0pt]
\item observe the \gls{cca} of each flow for a certain measurement period and
\item increase or decrease the buffer size for this flow depending on whether the scenario of \autoref{fig:tooLittle} or \autoref{fig:tooMuch} is true. 
\end{enumerate}
The challenge about this is to define the measurement period: As \autoref{fig:tooLittle} and \autoref{fig:tooMuch} show, we would like to use the measurement period that is the longest period between two packet losses. If this period is observed, it is possible to compute the standing queue as simply the minimum queue observed in this interval. For example, if the minimum queue is 3 packets it means, that there is a standing queue of 3 packets and the algorithm will reduce buffer size by 3 to eliminate the unnecessary queuing. Conversely, if the buffer is too small, the algorithm computes how long the link was underutilized and how many packets could have been transmitted during the idle period. The buffer size is then increased by this number of packets. 

This is not trivial because of the following consideration: If the measurement period is mistakenly not the longest interval shown in the figure, but the interval between two adjacent packet losses, the algorithm would assume an enormous standing queue and drastically reduce the buffer to be close to zero. This would lead to severe underutilization. 

The solution is thus to use a \textit{\gls{gi}}: The algorithm always waits for a specified minimum amount of time, picks the longest interval without packet loss in this guard interval and then applies its logic to reduce the standing queue on this \textit{\gls{li}}. We dynamically compute the \gls{gi} as a multiple or a fraction (parameter of our algorithm) of the previous \gls{li} (\autoref{fig:intervals}): 

\begin{figure}[h]
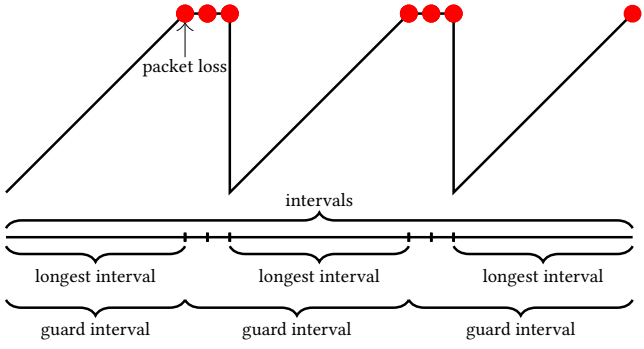

\includegraphics[width=\columnwidth]{{{figures/cocoa_intervals}}}
\caption{Illustration of the interval mechanism used by \gls{cocoa} with a multiplier of $\frac{1}{2}$. Each \textit{guard interval} (GI) contains one or more intervals of which the longest one is considered the \textit{longest interval} (LI). The \gls{gi} is always at least the length of the previous \gls{li} times the multiplier. A \gls{gi} ends at the first packet loss after the previous \gls{li} times the multiplier.}
\label{fig:intervals}
\end{figure}

For example, we choose the multiplier to be $\frac{1}{2}$. Now, if the \gls{li} in the previous \gls{gi} was 100\,ms, then the next \gls{gi} is at least $\frac{1}{2} 100\,ms$. During the \gls{gi} we monitor all intervals and if the current interval is the longest one, it becomes the new \gls{li}. The \gls{gi} ends when the first packet loss occurs after the minimum duration of the \gls{gi} elapsed. 

\begin{algorithm}[h]
\caption{Procedure that is executed when a new packet is received to be enqueued.}
\label{alg:enqueue}
\begin{algorithmic}[1]%
\Function{enqueue}{new\_packet}
	\If{queue is full}
		\If{link was idle during this interval $\land$ the buffer wasn't already enlarged in this interval $\land$ this is not the first interval of this flow}
			\State buffer\_size $\gets$ buffer\_size + (packets\_transmitted\_in\_current\_interval / time\_active $\times$ time\_idle)
			\State add new\_packet to queue
		\ElsIf{the buffer was enlarged in this interval $\lor$ this is the first interval of this flow}
			\State start a new interval
			\State start a new \gls{gi}
			\State drop new\_packet
		\ElsIf{current\_time $\geq$ end\_the\_current\_\gls{gi}}
			\If{there was a standing queue in the \gls{li} of the \gls{gi}}
				\State buffer\_size $\gets$ buffer\_size - standing\_queue
				\State drop superfluous packets
			\EndIf
			\State start a new interval
			\State start a new \gls{gi}
			\State drop new\_packet
		\Else
			\State start a new interval
			\State drop new\_packet
		\EndIf
	\Else
		\State add new\_packet to queue
	\EndIf
\EndFunction
\end{algorithmic}
\end{algorithm}
Almost all of the logic of our algorithm is executed when a new packet is received to be enqueued. The full algorithm is depicted as \autoref{alg:enqueue}.

\subsection{Choice of the multiplier}

The purpose of the multiplier, which is used to multiply the \gls{li} of previous \gls{gi} to define the new \gls{gi}, is to prevent that the short gap between two subsequent packet losses is used as the interval to determine whether the buffer size has to be adapted. This means that the multiplier must be large enough so that it allows to skip over the packet losses that occur when the buffer is full. For Reno and Cubic, the loss-free period is very large compared to the period during which packet loss occurs. Specifically, the period with packet loss has the length of one \gls{rtt} since Reno and Cubic reduce their sending rate as soon as packet loss occurs, thus ending the packet loss. This means that the multiplier can be very small in order to work for these \glspl{cca}. We conducted experiments with the multiplier being $0.5$ and have never encountered the problem that the interval between adjacent packet losses is considered the longest interval and that the buffer is cut to virtually nothing because a standing queue is assumed. However, BBR~v1 uses a completely different approach compared to Reno and Cubic to probe for bandwidth: It periodically increases its sending rate and then reduces it again to see if it can achieve more throughput. This happens with the following pattern: $[\frac{5}{4},\frac{3}{4},1,1,1,1,1,1]$. The sending rate is increased to probe for more bandwidth, then decreased to reduce the standing queue potentially formed and after that BBR is simply keeping the sending rate to match the estimated bandwidth. Each of these eight phases lasts one \gls{rtt}. BBR uses a random cyclic permutation of this pattern with the only condition being that it cannot start with $\frac{3}{4}$. The following pattern could thus occur: $[1,1,1,1,1,1,\frac{5}{4},\frac{3}{4},\frac{5}{4},\frac{3}{4},1,1,1,1,1,1]$. If the \gls{gi} ends after the first 6 phases and also the \gls{li} was these six phase, with a factor of $0.5$, the next \gls{gi} would thus end at the end of phase 8. Then, the \gls{li} would be the 8th cycle. The next \gls{gi} would then be of the duration of $0.5$ phases and would then end in the middle of cycle 9. This would be erroneous because this is a probing phase during which packet loss occurs. The aforementioned problem of erroneously and significantly reducing the buffer could thus occur. This means that for BBR~v1 the multiplier must be larger than 1 since it can happen that non-probing phases are followed by probing phases of the same length. From these considerations it follows that for our experiments we generally use a multiplier of $1.25$. The behavior is different for BBR~v2 which drastically changed the probing phase and now consists of a random 2-3\,s phase with a constant sending rate (cruising) followed by a quick probing phase followed by a decrease phase \cite{cardwell_bbr_2019}. We thus conjecture that a multiplier of $0.5$ is sufficient for BBR~v2. However, we could not finally confirm this since BBR~v2 is still under development. 

\subsection{Maximum buffer increase}

Another potential problem is the following: If the available share of bandwidth suddenly increases drastically, our algorithm would sharply increase the buffer size. In this case it can happen that the buffer becomes much too large. We thus limit the maximum buffer increase to be 2 times the previous buffer size. This parameter is only relevant in the case of a sudden increase of available bandwidth or if a flow's throughput is application limited and not during stable state behavior. 

\subsection{Maximum \gls{gi}}

One more challenge is posed by the following consideration: If a delay-based \gls{cca} is in its stable state, no packet loss should occur. The \gls{li} could thus become very large (like tens of seconds) and also the next \gls{gi} would be extremely large. This can be a problem if a delay-based \gls{cca} like Vegas \cite{brakmo_tcp_1995} is used for a long flow and then after several seconds, the bandwidth drastically lowers. Then suddenly packet loss could occur but the \gls{gi} would be so long that nothing would happen for a long time. To counter this, we add a parameter that specifies the maximum duration of the \gls{gi}. We set this parameter to 1\,s. This means that we also assume that no flows are handled by \gls{cocoa} which have an \gls{rtt} larger than 1\,s. Another possibility for setting this parameter is to use a multiple of the \gls{rtt}, for example 2, 4 or 8 times the most recent \gls{rtt}. \gls{rtt} could be measured using TCP timestamps \cite{borman_tcp_2014} in case of TCP flows or the spin bit \cite{kuhlewind_quic_2018} for QUIC. 

\section{Implementation}

We implement \gls{cocoa} as an extension of the ``fq'' \gls{qdisc} \cite{dumazet_pkt_sched:_2013}. This means that when using \gls{cocoa} also all features offered by fq are available. We add three configuration parameters to the \gls{qdisc}: the multiplier (default 1.25), the maximum buffer increase (default 2.0) and the maximum \gls{gi} duration (default 1.0\,s). We make the source code of our implementation freely available to enable reproducibility and encourage experimentation: \url{https://github.com/CN-TU/cocoa-qdisc}.

\section{Evaluation}

We evaluate \gls{cocoa} using the \texttt{py-virtnet} (\url{https://github.com/CN-TU/py-virtnet}) toolkit to build a virtual network using Linux's network namespaces. We initialize the buffer size for \gls{cocoa} to 100 packets per flow like in fq. The testbed consists of a sender and a receiver connected via a switch. We run the \gls{qdisc} being tested on the interface that connects the switch to the receiver. We introduce delay with \texttt{netem} and limit bandwidth with \texttt{htb}. 

\begin{figure}[h]
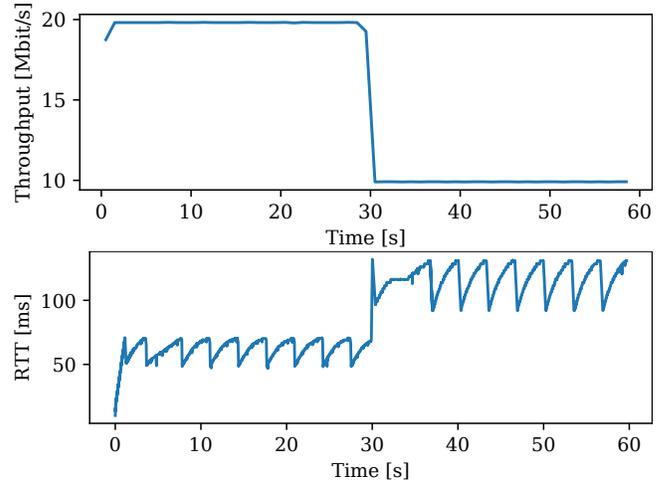

\includegraphics[width=\columnwidth]{{{plots/throughput_1_fq_cubic_10_20_60_0.5_bw_1571240937211.pcap}}}
\includegraphics[width=\columnwidth]{{{plots/rtt_1_fq_cubic_10_20_60_0.5_bw_1571240937211.pcap}}}
\caption{A Cubic flow with fq at the bottleneck. The initial bandwidth is 20\,Mbit/s but we halve it after 30\,s. The delay is 10\,ms. It is clear that there is a standing queue and after the bandwidth halves, the minimum delay is 100\,ms even though 10\,ms would be possible.}
\label{fig:fq}
\end{figure}

\begin{figure}[h]
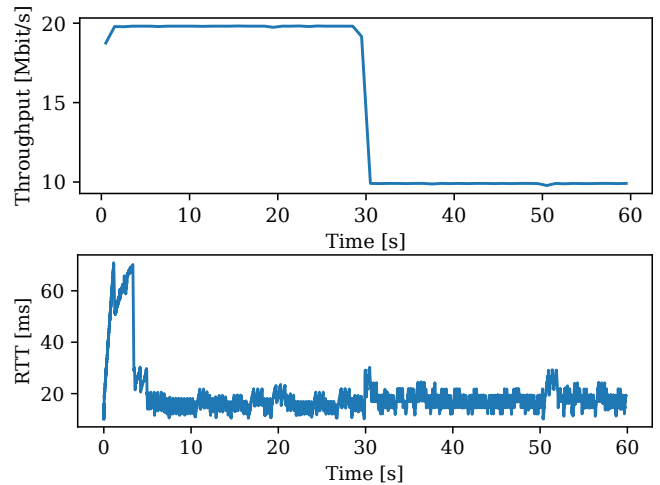

\includegraphics[width=\columnwidth]{{{plots/throughput_1_cn_cubic_10_20_60_0.5_bw_1571241047054.pcap}}}
\includegraphics[width=\columnwidth]{{{plots/rtt_1_cn_cubic_10_20_60_0.5_bw_1571241047054.pcap}}}
\caption{A Cubic flow with \gls{cocoa} at the bottleneck. The initial bandwidth is 20\,Mbit/s but we halve it after 30\,s. The delay is 10\,ms. \gls{cocoa} keeps the queue smaller than fq while achieving the same throughput.}
\label{fig:cn}
\end{figure}

\begin{figure}[h]
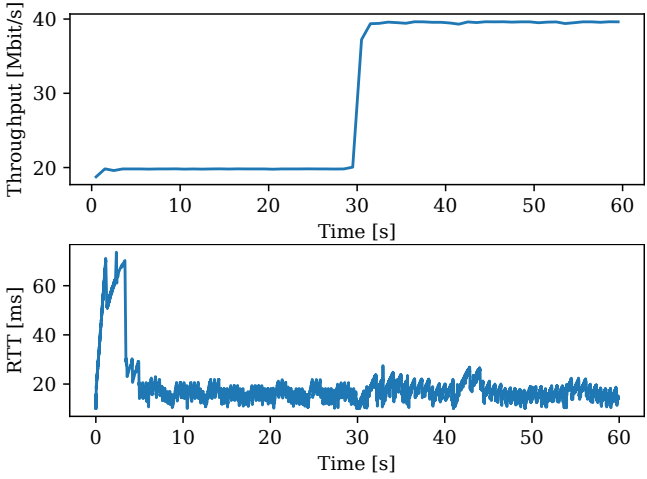

\includegraphics[width=\columnwidth]{{{plots/throughput_1_cn_cubic_10_20_60_2.0_bw_1571241151290.pcap}}}
\includegraphics[width=\columnwidth]{{{plots/rtt_1_cn_cubic_10_20_60_2.0_bw_1571241151290.pcap}}}
\caption{A Cubic flow with \gls{cocoa} at the bottleneck. The initial bandwidth is 20\,Mbit/s but we double it after 30\,s. The delay is 10\,ms. Also a sudden large increase of bandwidth is handled well by \gls{cocoa}.}
\label{fig:cn2}
\end{figure}

First we evaluate if \gls{cocoa} is able to maintain a small buffer like it is necessary in case of small \glspl{bdp}. As can be seen in \autoref{fig:fq} the regular fq qdisc (with a standard queue size of 100 packets) maintains a standing queue and unnecessarily leads to a significant increase in \gls{rtt} (100\,ms minimum, when the real minimum is 10\,ms). In contrast, \autoref{fig:cn} maintains full throughput while at the same time keeping the delay minimal and not having a standing queue. Also, when drastically changing the bandwidth by halving it (\autoref{fig:cn}) or doubling it (\autoref{fig:cn2}) \gls{cocoa} rapidly adapts to the new conditions and returns to the optimum state. 

\begin{figure}[h]
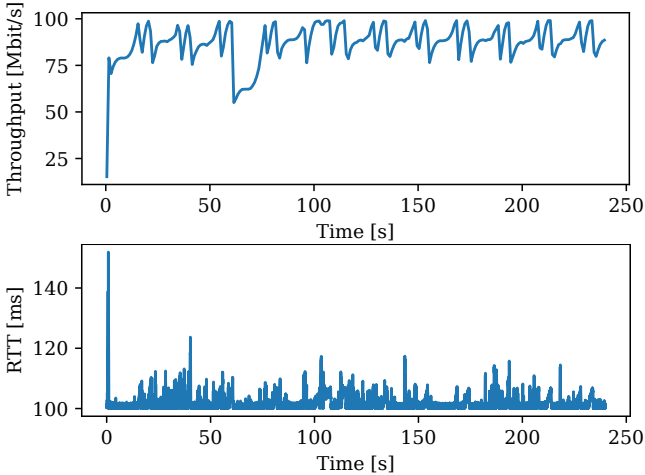

\includegraphics[width=\columnwidth]{{{plots/throughput_1_fq_codel_cubic_100_100_240_1.0_bw_1571244312838.pcap}}}
\includegraphics[width=\columnwidth]{{{plots/rtt_1_fq_codel_cubic_100_100_240_1.0_bw_1571244312838.pcap}}}
\caption{A Cubic flow with fq\_codel at the bottleneck. The bandwidth is 100\,Mbit/s. The delay is 100\,ms. The average \gls{rtt} is 101.61\,ms. Total throughput is 2615\,MB.}
\label{fig:fqCodel}
\end{figure}

\begin{figure}[h]
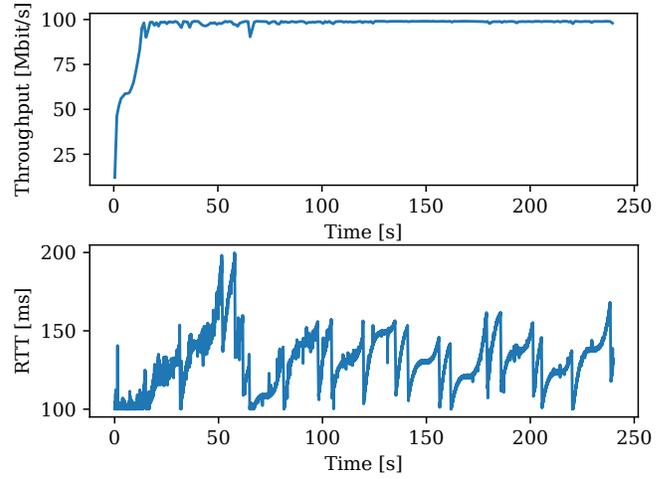

\includegraphics[width=\columnwidth]{{{plots/throughput_1_cn_cubic_100_100_240_1.0_bw_1571244036720.pcap}}}
\includegraphics[width=\columnwidth]{{{plots/rtt_1_cn_cubic_100_100_240_1.0_bw_1571244036720.pcap}}}
\caption{A Cubic flow with \gls{cocoa} at the bottleneck. The bandwidth is 100\,Mbit/s. The delay is 100\,ms. The average \gls{rtt} is 130.36\,ms. Total throughput is 2893\,MB.}
\label{fig:cnLong}
\end{figure}

\begin{table}[t]
\caption{Comparison of \glspl{qdisc} for loss-based \glspl{cca} on a link of 100\;Mbit/s with an \gls{rtt} of 50\;ms. Utilization in \% of maximum capacity; \gls{rtt} in milliseconds. For Cubic fq achieves good performance. This is because its default buffer size of 100 packets coincidentially works well for this specific link speed and \gls{rtt}. We take the mean of 10 runs for each scenario.} \label{tab:performance_results}
\begin{tabular}{l r r r r r r} \toprule
& \multicolumn{2}{l}{fq} & \multicolumn{2}{l}{fq\_codel} & \multicolumn{2}{l}{\gls{cocoa}} \\
\cmidrule(r){2-3} \cmidrule(lr){4-5} \cmidrule(l){6-7}
& Util. & \gls{rtt} & Util. & \gls{rtt} & Util. & \gls{rtt} \\ \midrule
Cubic	& 96.2 & 54 & 92.6 & 52 & 98.5 & 66.5 \\
Reno	& 84.1 & 52.7 & 81.1 & 51.5 & 97.9 & 98.2 \\
\bottomrule
\end{tabular}
\end{table}

Next we compare \gls{cocoa} against fq\_codel. \autoref{fig:fqCodel} shows that fq\_codel keeps the queuing delay under 5\,ms. It also shows that keeping the queuing delay that low is detrimental for achieving full throughput when using the Cubic \gls{cca}. With \gls{cocoa} we achieve throughput that is more than 10\% higher overall, while only slightly increasing the average delay. In addition to Cubic, we performed experiments with Reno. Here, over a 240\,s flow of 100\,Mbit/s with a delay of 50\,ms, \gls{cocoa} achieves more than 20\% higher throughput than fq\_codel. The average \gls{rtt} for fq\_codel is 51.34\,ms while it is 96.78\,ms for \gls{cocoa}. \autoref{tab:performance_results} shows that \gls{cocoa} reaches link utilizations close to 100\% for both Cubic and Reno in large \gls{bdp} scenarios. \gls{rtt} can be higher than for other \glspl{qdisc} if this is required to achieve optimum throughput for the \gls{cca} for the given link speed and base \gls{rtt}. 

Besides fq\_codel possibly leading to lower throughput as we have shown, another problem is that for links with a small \gls{bdp} fq\_codel can keep a standing queue akin to fq since fq\_codel only wants queuing delay to fall below 5\,ms once every 100\,ms. For example, on a link with a base \gls{rtt} of 1\,ms, fq\_codel would accept a permanent standing queue of 4.9\,ms, leading to overall \gls{rtt} being more than doubled, which is not required for achieving optimal throughput. 

We also ran experiments with the cake \gls{qdisc}. However, results were very similar to those of fq\_codel, which is no surprise since cake is based on fq\_codel.

\begin{figure}[h]
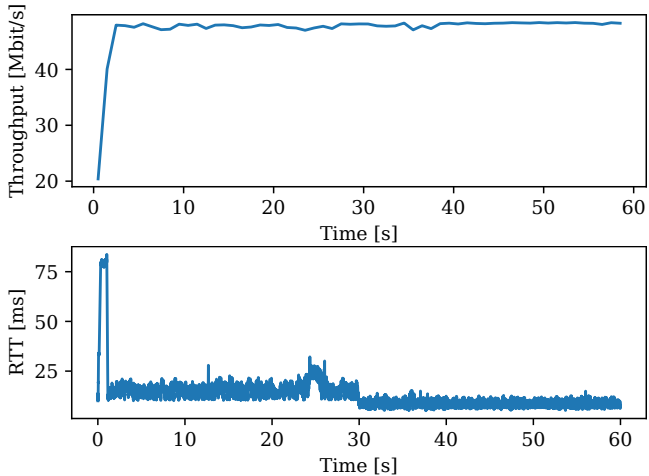

\includegraphics[width=\columnwidth]{{{plots/throughput_1_cn_bbr_10_50_60_0.5_delay_1571288916582.pcap}}}
\includegraphics[width=\columnwidth]{{{plots/rtt_1_cn_bbr_10_50_60_0.5_delay_1571288916582.pcap}}}
\caption{A BBR~v1 flow with \gls{cocoa} at the bottleneck. The bandwidth is 50\,Mbit/s. The delay is 10\,ms and we halve it after 30\,s.}
\label{fig:cnBBR}
\end{figure}

Also for BBR~v1, throughput quickly reaches its maximum while no standing queue is forming (\autoref{fig:cnBBR}). 

\section{Discussion}

Our goal was to design a fair \gls{qdisc} which achieves maximum throughput while keeping the queuing delay as small as possible. The results of the experiments, which we perform with a prototype of \gls{cocoa}, show that there are certain common scenarios in which current state-of-the-art fair \glspl{qdisc} like fq, fq\_codel and cake fall short of the optimum throughput by a significant margin while our approach succeeds in fully utilizing the bottleneck link (with a potential increase in delay) (\autoref{tab:performance_results}; \autoref{fig:fqCodel} vs.~\autoref{fig:cnLong}). Furthermore, the aforementioned \glspl{qdisc} also suffer from standing queues in scenarios with small \gls{bdp} which we can also mitigate with our approach (\autoref{fig:fq} vs.~\autoref{fig:cn}). 

We see this work as an initial step towards learning and flow-adaptive fair queuing at bottlenecks and think that the most promising application domain of \gls{cocoa} is at the Internet's edge. An interesting further improvement could be to not statically initialize the buffer with a constant 100 packets but instead use experience from previous flows. Moreover, we think that a promising direction for future work might be to explore the use of reinforcement learning. Such an approach would fingerprint a flow and choose the buffer size accordingly, maximizing a chosen objective such as high throughput with minimal delay. 

\section*{Acknowledgements}
We thank Gernot Vormayr for providing us with \texttt{py-virtnet}, a toolkit for easily building virtual networks. 

\bibliographystyle{ACM-Reference-Format}
\bibliography{bibliography}


\begin{thebibliography}{13}


\ifx \showCODEN    \undefined \def \showCODEN     #1{\unskip}     \fi
\ifx \showDOI      \undefined \def \showDOI       #1{#1}\fi
\ifx \showISBNx    \undefined \def \showISBNx     #1{\unskip}     \fi
\ifx \showISBNxiii \undefined \def \showISBNxiii  #1{\unskip}     \fi
\ifx \showISSN     \undefined \def \showISSN      #1{\unskip}     \fi
\ifx \showLCCN     \undefined \def \showLCCN      #1{\unskip}     \fi
\ifx \shownote     \undefined \def \shownote      #1{#1}          \fi
\ifx \showarticletitle \undefined \def \showarticletitle #1{#1}   \fi
\ifx \showURL      \undefined \def \showURL       {\relax}        \fi
\providecommand\bibfield[2]{#2}
\providecommand\bibinfo[2]{#2}
\providecommand\natexlab[1]{#1}
\providecommand\showeprint[2][]{arXiv:#2}

\bibitem[\protect\citeauthoryear{Bless, Hock, and Zitterbart}{Bless
  et~al\mbox{.}}{2018}]%
        {bless_policy-oriented_2018}
\bibfield{author}{\bibinfo{person}{Roland Bless}, \bibinfo{person}{Mario Hock},
  {and} \bibinfo{person}{Martina Zitterbart}.} \bibinfo{year}{2018}\natexlab{}.
\newblock \showarticletitle{Policy-oriented {AQM} {Steering}}. In
  \bibinfo{booktitle}{\emph{2018 {IFIP} {Networking} {Conference} ({IFIP}
  {Networking}) and {Workshops}}}. \bibinfo{pages}{1--9}.
\newblock


\bibitem[\protect\citeauthoryear{Borman, Scheffenegger, and Jacobson}{Borman
  et~al\mbox{.}}{2014}]%
        {borman_tcp_2014}
\bibfield{author}{\bibinfo{person}{David Borman}, \bibinfo{person}{Richard
  Scheffenegger}, {and} \bibinfo{person}{Van Jacobson}.}
  \bibinfo{year}{2014}\natexlab{}.
\newblock \bibinfo{title}{{TCP} {Extensions} for {High} {Performance}}.
\newblock
\newblock
\urldef\tempurl%
\url{https://tools.ietf.org/html/rfc7323}
\showURL{%
\tempurl}


\bibitem[\protect\citeauthoryear{Brakmo and Peterson}{Brakmo and
  Peterson}{1995}]%
        {brakmo_tcp_1995}
\bibfield{author}{\bibinfo{person}{L.S. Brakmo} {and} \bibinfo{person}{L.L.
  Peterson}.} \bibinfo{year}{1995}\natexlab{}.
\newblock \showarticletitle{{TCP} {Vegas}: end to end congestion avoidance on a
  global {Internet}}.
\newblock \bibinfo{journal}{\emph{IEEE Journal on Selected Areas in
  Communications}} \bibinfo{volume}{13}, \bibinfo{number}{8}
  (\bibinfo{date}{Oct.} \bibinfo{year}{1995}), \bibinfo{pages}{1465--1480}.
\newblock
\showISSN{0733-8716, 1558-0008}


\bibitem[\protect\citeauthoryear{Cardwell, Cheng, Gunn, Yeganeh, and
  Jacobson}{Cardwell et~al\mbox{.}}{2016}]%
        {cardwell_bbr:_2016}
\bibfield{author}{\bibinfo{person}{Neal Cardwell}, \bibinfo{person}{Yuchung
  Cheng}, \bibinfo{person}{C.~Stephen Gunn}, \bibinfo{person}{Soheil~Hassas
  Yeganeh}, {and} \bibinfo{person}{Van Jacobson}.}
  \bibinfo{year}{2016}\natexlab{}.
\newblock \showarticletitle{{BBR}: {Congestion}-{Based} {Congestion}
  {Control}}.
\newblock \bibinfo{journal}{\emph{ACM Queue}}  \bibinfo{volume}{14,
  September-October} (\bibinfo{year}{2016}), \bibinfo{pages}{20 -- 53}.
\newblock


\bibitem[\protect\citeauthoryear{Cardwell, Cheng, Yeganeh, Jha, Seung, Swett,
  Vasiliev, Wu, Mathis, and Jacobson}{Cardwell et~al\mbox{.}}{2019}]%
        {cardwell_bbr_2019}
\bibfield{author}{\bibinfo{person}{Neal Cardwell}, \bibinfo{person}{Yuchung
  Cheng}, \bibinfo{person}{Soheil~Hassas Yeganeh}, \bibinfo{person}{Priyaranjan
  Jha}, \bibinfo{person}{Yousuk Seung}, \bibinfo{person}{Ian Swett},
  \bibinfo{person}{Victor Vasiliev}, \bibinfo{person}{Bin Wu},
  \bibinfo{person}{Matt Mathis}, {and} \bibinfo{person}{Van Jacobson}.}
  \bibinfo{year}{2019}\natexlab{}.
\newblock \showarticletitle{{BBR} v2: {A} {Model}-based {Congestion} {Control}
  {IETF} 105 {Update}}.
\newblock  (\bibinfo{year}{2019}), \bibinfo{pages}{21}.
\newblock


\bibitem[\protect\citeauthoryear{Dumazet}{Dumazet}{2013}]%
        {dumazet_pkt_sched:_2013}
\bibfield{author}{\bibinfo{person}{Eric Dumazet}.}
  \bibinfo{year}{2013}\natexlab{}.
\newblock \bibinfo{title}{pkt\_sched: fq: {Fair} {Queue} packet scheduler
  [{LWN}.net]}.
\newblock
\newblock
\urldef\tempurl%
\url{https://lwn.net/Articles/565421/}
\showURL{%
\tempurl}


\bibitem[\protect\citeauthoryear{Ha, Rhee, and Xu}{Ha et~al\mbox{.}}{2008}]%
        {ha_cubic:_2008}
\bibfield{author}{\bibinfo{person}{Sangtae Ha}, \bibinfo{person}{Injong Rhee},
  {and} \bibinfo{person}{Lisong Xu}.} \bibinfo{year}{2008}\natexlab{}.
\newblock \showarticletitle{{CUBIC}: a new {TCP}-friendly high-speed {TCP}
  variant}.
\newblock \bibinfo{journal}{\emph{ACM SIGOPS Operating Systems Review}}
  \bibinfo{volume}{42}, \bibinfo{number}{5} (\bibinfo{date}{July}
  \bibinfo{year}{2008}), \bibinfo{pages}{64--74}.
\newblock
\showISSN{01635980}


\bibitem[\protect\citeauthoryear{Høiland-Jørgensen, Täht, and
  Morton}{Høiland-Jørgensen et~al\mbox{.}}{2018}]%
        {hoiland-jorgensen_piece_2018}
\bibfield{author}{\bibinfo{person}{T. Høiland-Jørgensen}, \bibinfo{person}{D.
  Täht}, {and} \bibinfo{person}{J. Morton}.} \bibinfo{year}{2018}\natexlab{}.
\newblock \showarticletitle{Piece of {CAKE}: {A} {Comprehensive} {Queue}
  {Management} {Solution} for {Home} {Gateways}}. In
  \bibinfo{booktitle}{\emph{2018 {IEEE} {International} {Symposium} on {Local}
  and {Metropolitan} {Area} {Networks} ({LANMAN})}}. \bibinfo{pages}{37--42}.
\newblock


\bibitem[\protect\citeauthoryear{Jacobson}{Jacobson}{1988}]%
        {jacobson_congestion_1988}
\bibfield{author}{\bibinfo{person}{V. Jacobson}.}
  \bibinfo{year}{1988}\natexlab{}.
\newblock \showarticletitle{Congestion {Avoidance} and {Control}}. In
  \bibinfo{booktitle}{\emph{Symposium {Proceedings} on {Communications}
  {Architectures} and {Protocols}}} \emph{(\bibinfo{series}{{SIGCOMM} '88})}.
  \bibinfo{publisher}{ACM}, \bibinfo{address}{New York, NY, USA},
  \bibinfo{pages}{314--329}.
\newblock
\showISBNx{978-0-89791-279-2}


\bibitem[\protect\citeauthoryear{Kühlewind and Trammell}{Kühlewind and
  Trammell}{2018}]%
        {kuhlewind_quic_2018}
\bibfield{author}{\bibinfo{person}{Mirja Kühlewind} {and}
  \bibinfo{person}{Brian Trammell}.} \bibinfo{year}{2018}\natexlab{}.
\newblock \bibinfo{title}{The {QUIC} {Latency} {Spin} {Bit}}.
\newblock
\newblock
\urldef\tempurl%
\url{https://tools.ietf.org/html/draft-ietf-quic-spin-exp-01}
\showURL{%
\tempurl}


\bibitem[\protect\citeauthoryear{Nichols and Jacobson}{Nichols and
  Jacobson}{2012}]%
        {nichols_controlling_2012}
\bibfield{author}{\bibinfo{person}{Kathleen Nichols} {and} \bibinfo{person}{Van
  Jacobson}.} \bibinfo{year}{2012}\natexlab{}.
\newblock \showarticletitle{Controlling {Queue} {Delay}}.
\newblock \bibinfo{journal}{\emph{Queue}} \bibinfo{volume}{10},
  \bibinfo{number}{5} (\bibinfo{date}{May} \bibinfo{year}{2012}),
  \bibinfo{pages}{20:20--20:34}.
\newblock
\showISSN{1542-7730}


\bibitem[\protect\citeauthoryear{Shreedhar and Varghese}{Shreedhar and
  Varghese}{1996}]%
        {shreedhar_efficient_1996}
\bibfield{author}{\bibinfo{person}{M. Shreedhar} {and} \bibinfo{person}{G.
  Varghese}.} \bibinfo{year}{1996}\natexlab{}.
\newblock \showarticletitle{Efficient fair queuing using deficit round-robin}.
\newblock \bibinfo{journal}{\emph{IEEE/ACM Transactions on Networking}}
  \bibinfo{volume}{4}, \bibinfo{number}{3} (\bibinfo{date}{June}
  \bibinfo{year}{1996}), \bibinfo{pages}{375--385}.
\newblock


\bibitem[\protect\citeauthoryear{Taht, Gettys, Hoeiland-Joergensen,
  Hoeiland-Joergensen, Dumazet, Gettys, Dumazet, and McKenney}{Taht
  et~al\mbox{.}}{2018}]%
        {taht_flow_2018}
\bibfield{author}{\bibinfo{person}{D. Taht}, \bibinfo{person}{Jim Gettys},
  \bibinfo{person}{T. Hoeiland-Joergensen}, \bibinfo{person}{Toke
  Hoeiland-Joergensen}, \bibinfo{person}{Eric Dumazet}, \bibinfo{person}{J.
  Gettys}, \bibinfo{person}{E. Dumazet}, {and} \bibinfo{person}{P. McKenney}.}
  \bibinfo{year}{2018}\natexlab{}.
\newblock \bibinfo{title}{The {Flow} {Queue} {CoDel} {Packet} {Scheduler} and
  {Active} {Queue} {Management} {Algorithm}}.
\newblock
\newblock
\urldef\tempurl%
\url{https://tools.ietf.org/html/rfc8290}
\showURL{%
\tempurl}


\end{thebibliography}

\end{document}